\documentclass[11pt]{article}
\usepackage[latin1]{inputenc}
\usepackage[english]{babel}
\usepackage[namelimits]{amsmath}
\usepackage{amssymb}
\usepackage{amsmath}
\usepackage{amsthm}

\begin{document}
\title{MODELING USUAL AND UNUSUAL ANISOTROPIC SPHERES}
\author{Stefano Viaggiu
\\
Dipartimento di Matematica, Universit\'a "Tor Vergata"',\\ 
Via della Ricerca Scientifica, 1\\
Rome, Italy 00133,\\
viaggiu@axp.mat.uniroma2.it}
\date{\today}\maketitle
\begin{abstract}
In this paper, we study anisotropic spheres built from 
known static spherical solutions. 
In particular, we are interested in the physical
consequences of a "small" departure from a physically sensible
configuration.
The obtained solutions smoothly depend on free parameters. 
By setting these parameters to zero, the starting seed solution is
regained. We apply our
procedure in detail by taking as seed solutions the Florides metrics, 
and the Tolman IV solution. We show that the chosen Tolman IV,
and also Heint IIa Durg IV,V
perfect fluid solutions,
can be used to generate a class of parametric solutions where the 
anisotropic factor has 
features recalling boson stars. This is an indication that boson 
stars could emerge by "perturbing" appropriately a perfect fluid solution
(at least for the seed metrics considered). Finally,
starting with Tolman IV, Heint IIa and Durg IV,V
solutions, we build anisotropic gravastar-like 
sources
with
the appropriate boundary conditions.  
\end{abstract}
\section{Introduction}
In the real world, stars are not made up of perfect fluids. Despite this, it
seems that perfect fluids are appropriate for describing ordinary astrophysical 
objects. However, since of the work of Bower and Liang \cite{1}, anisotropic 
models have been attracting increasing interest \cite{2}. Firstly, anisotropic fluids
have been considered in studying objects with hight redshift such as quasars \cite{3}.
Local anisotropies can arise from modeling very compact objects such as 
neutron stars.
Further, anisotropies can lead to important modifications of the physical
parameters of the fluid 
\cite{4,5,6}.
In particular, these can affect the critical mass and the stability of stars
\cite{7,8} and also the redshift. In this context, particularly 
interesting is the 
case of boson stars {\cite{9,14}, where naturally anisotropies come in action.
These are hypothetical stars, born of an idea of Kaup \cite{15}, constituting
a bosonic particle solution of the Klein-Gordon equation coupled with 
general relativity. Bosons are described with a complex scalar field 
${\Phi}(r,t)$,: ${\Phi}(r,t)=J(r)e^{-\imath\sigma t}$, with $\sigma$
the frequency and $(r,t)$ the usual coordinates in a static spherical space-time.
The function $J(r)$ is a measure of the anisotropy of the star. These stars 
always have $(P_{\perp}-P_{r})\leq 0 $, with $P_{\perp}$ and $P_{r}$ respectively
the tangential and the radial pressure, and $(P_{\perp}-P_{r})$ 
vanishes at the centre
$r=0$ and near the approximate radius of 
the star (see \cite{11} and references therein).  
From the considerations above, 
it is evident
that there is a need to investigate anisotropies in general relativity. 
In particular, it is 
useful to build anisotropic metrics that differ from a given seed metric by 
free parameters.
In the literature \cite{1,11,16,17}, methods have been developed to build 
explicit anisotropic 
spheres. 
These methods require, before solving the Tolman equation
\cite{18},
an ansatz regarding the expression of the anisotropy
$(P_{\perp}-P_{r})$. Generally, only solutions with constant density $E$ or with
$E\propto\frac{1}{r^2}$ are pratically available \cite{1,11}. In other cases
\cite{16,17}, anisotropy factor is calculated by means of an 
integration involving
the chosen seed perfect fluid configuration. Also in this way, only particular
configurations are available. 
Recently \cite{a1}, all the anisotropic static spherical solutions have been 
obtained via quadratures
by specifying two generating functions .\\
Motivated by these facts, we derive a simple procedure for obtaining workable 
regular 
anisotropic metrics with the anisotropy factor depending on arbitrary regular
functions satisfying some boundary conditions.
We point out that  this paper is not principally
devoted to the generating
method, but rather
to the physical
consequences of a departure from a physically sensible
configuration. In particular, we are interested in possible
phase transitions of the matter caused by a "small" departure from
a given starting space-time. We show as a "dynamic" describing
this deparure can be introduced.
We present our procedure in detail with 
two seed space-times: the Florides \cite{19} class of solutions with
vanishing radial pressure and a particular Tolman \cite{20} perfect fluid
solution. Further, we show that the technique can also be applied, at least,
to other perfect
fluid seed metrics than the Tolman IV one, i.e. Heint IIa, Durg IV and Durg V.
In section 2 we describe our simple procedure together with the regularity
conditions on the metric coefficients. In section 3 the technique is applied
to the Florides solutions and in section 4 to the Tolman IV solution and to 
the Heint IIa metric. 
In section 5, starting with the Tolman IV and Heint IIa 
metrics, we build a simple gravastar
model. Gravastar models can also be built (at least)
starting with Durg IV, V
spacetimes.
Section 6 presents some final remarks and conclusions.

\section{General Facts}   
Our starting point is the line element appropriate for static spheres:
\begin{equation}
ds^2=A(r)dr^2+r^2[d{\theta}^2+{\sin}^2\theta d{\phi}^2]-B(r)dt^2.
\label{1}
\end{equation}
The most general expression for the energy-momentum tensor $T_{ab}$ is:
\begin{equation}
T_{ab}=E(r)V_aV_b+P_{\perp}(r)[W_aW_b+L_aL_b]+P_{r}(r)S_aS_b,
\label{2}
\end{equation}
where $E(r)$ is the energy-density, $P_{\perp}(r)$ the tangential pressure and 
$P_r(r)$ the radial pressure, with
(Einstein's equations are $G_{ab}=-T_{ab}$)
\begin{eqnarray}
& &V_a = \left[0, 0, 0, -\sqrt{B(r)}\right],\label{3}\\
& &W_a = \left[0, r, 0, 0\right],\nonumber\\
& &L_a = \left[0, 0, r \sin\theta, 0\right],\nonumber\\
& &S_a = \left[\sqrt{A(r)}, 0, 0, 0\right].\nonumber\\
\end{eqnarray} 
In terms of $A(r),B(r)$, the Einstein's equations reduce to:
\begin{eqnarray}
& &E=\frac{rA_{,r}+A^2-A}{r^2A^2},\label{4}\\
& &P_r=\frac{1}{r^2A}\left[1-A+r\frac{B_{,r}}{B}\right],\label{5}\\
& &P_{\perp}=\frac{1}{4rA^2B^2}X(r),\label{6}\\
& &X(r)=2B_{,r}AB-2A_{,r}B^2+
2rB_{,r,r}AB-rA_{,r}B_{,r}B-rB_{,r}^2A,\nonumber\\
& &P_{\perp}-P_{r}=S(r)=-\frac{1}{4r^2A^2B^2}\Gamma(r),\label{7}\\
& &\Gamma(r)=2rA_{,r}B^2+2rB_{,r}AB-2r^2B_{,r,r}AB+r^2B_{,r}A_{,r}B-
4A^2B^2+\nonumber\\
& &4AB^2+r^2AB_{,r}^{2},\nonumber
\end{eqnarray}  
where subindices with comma denote partial derivative.\\
Now, suppose that $A(r), B(r)$ 
depend smoothly on certain parameters
$a_i,b_i$ respectively
in such a way that, setting $a_i=b_i=0$,
we have
\begin{equation}
A(r)=A_0(r)\;\;,\;\;B(r)=B_{0}(r),
\label{9}
\end{equation}
$A_0, B_0$ being given seed solutions.\\
Under the assumption that the seed solution is regular, satisfies all
energy conditions and is smoothly matched to the vacuum Schwarzschild 
solution, we show that the solution $A(r),B(r)$ also is, 
provided that $a_i,b_i$ is chosen appropriately and that, 
by inspection of (\ref{4})-(\ref{7}), the following 
conditions are fulfilled:
\begin{eqnarray}
& &A(r)>0\;\forall r \in [0,R]\;,\;A(R)=A_0(R),\label{10}\\
& &A(0)=1+\alpha r^n+o(1)\;,\;\alpha \in \mathbf{R}\;,\;n\geq 2,\nonumber\\
& &B(r)>0\;\forall r \in [0,R]\;,\;B(R)=B_0(R)\;,\;
B_{,r}(R)=B_{0,r}(R),\nonumber\\
& &B_{,r}(0)=\beta r^n+o(1)\;,\;\beta \in \mathbf{R}\;,\;n\geq 1,
\nonumber
\end{eqnarray}
$R$ being the radius of the star.
Further, if the seed metric is a perfect fluid solution or anisotropic matter
with $P_{0\perp}(R)=0$, then the generated solution will have $S(R)=0$, 
provided that
\begin{equation}
A_{,r}(R)=A_{0,r}(R)\;\;,\;\;B_{,r,r}(R)=B_{0,r,r}(R).
\label{11}
\end{equation}
Finally, all energy conditions and physical reasonability 
for usual ordinary matter require that
\begin{equation}
P_r\geq 0, P_{\perp}\geq 0, E\geq 0\;,\;
E-P_{\perp}\geq 0\;,\;E-P_r\geq 0\;\forall r \in [0,R].
\label{11b}
\end{equation}
For usual matter, all of the conditions (\ref{10})-(\ref{11b}) must
follow. Conversely, we can build solutions with non-usual 
matter, by dropping some of the conditions (\ref{11b}). 
In the next two sections, we show that the procedure described above
can effectively generate physically reasonable anisotropic fluids.\\
As seed metrics we take the Florides solution without radial pressure 
and with constant energy-density and the perfect fluid solutions
called Tolman IV and Heint IIa.

\section{Example 1: Florides seed solution}
As a first example, we take as seed metric the Florides \cite{19}
class of solutions representing anisotropic static spheres
with vanishing radial pressure. The most interesting feature
of these solutions is that the tangential pressure is sufficient
to prevent the gravitational collapse of a dustlike distribution
of matter. Further, these solutions describe the interior field of
Einstein's clusters where the tangential pressure is generated by a gas of
noncollisional particles moving on spheres. From (\ref{5}) we see that these
solutions have
\begin{eqnarray}
& &A_{0}(r)=\frac{1}{1-\frac{2m_{0}(r)}{r}}=
1+r\frac{B_{0,r}}{B_{0}},\label{12}\\
& &m_{0}(r)=\frac{1}{2}{\int}_{0}^{r}E_{0}(u)u^2 du\;\;,\;\;m_{0}(R)=M,\nonumber
\end{eqnarray}
where $m_{0}(r)$ is the gravitational mass inside a radius $r$.
As a title of 
example, we consider the solution in \cite{19} with constant 
energy-density given by
\begin{eqnarray}
& &A_{0}(r)=\frac{3}{3-r^2E}\;\;,\;\;E=\frac{6M}{R^3}\;\;,\;\;
P_{0\perp}(r)=\frac{E_{0}^2r^2}{4[3-r^2E_{0}]},\nonumber\\
& &P_{0r}=0\;\;,\;\;B_{0}(r)={\left(1-\frac{R^2 E}{3}\right)}^{\frac{3}{2}}
{\left(1-\frac{r^2 E}{3}\right)}^{\frac{-1}{2}}.\label{13}
\end{eqnarray}
We take
\begin{equation}
A=A_{0}\;,\;B=B_{0} e^{H(r)}\;,\;E=E_{0}. 
\label{14}
\end{equation}
The choice for $B(r)$ in (\ref{14}) is not restrictive and pemits
simpler computations. With (\ref{14}), for $P_r$ we read
\begin{equation}
P_r=\frac{H_{,r}}{rA}=\frac{H_{,r}}{3r}(3-r^2E).
\label{15}
\end{equation}
The expression (\ref{15}) implies that, in order to obtain physically
reasonable solutions, the function $H(r)$, adjoining
(\ref{10})-(\ref{11b}), must be monotonically increasing.\\
As an example, we choose
\begin{equation}
H=a{\left(\frac{r^2}{R^2}-1\right)}^3\;\;,\;\;a>0.
\label{16}
\end{equation}
With (\ref{16}) we get
\begin{equation}
P_r=\frac{2a}{R^6}\left(3-Er^2\right){\left(r^2-R^2\right)}^2.
\label{17}
\end{equation}
In this way, $P_r$ is regular, positive, vanishing at the boundary
$r=R$ and monotonically decreasing.
Therefore, energy conditions impose $E-P_r\geq 0$, i.e.
\begin{equation}
a\leq\frac{M}{R}=\frac{1}{k}.
\label{18}
\end{equation}
Regarding $P_{\perp}$, we have
\begin{eqnarray}
& &P_{\perp}=P_{0\perp}+\frac{6a}{R^9}
\left(R^2-r^2\right)\left(R^2-3r^2\right)
\left(R^3-2r^2M\right)+\nonumber\\
& &\frac{9a^2r^2}{R^{15}}{\left(R^2-r^2\right)}^4
\left(R^3-2r^2M\right).\label{19}
\end{eqnarray}
The first condition to be imposed for (\ref{19}) is that $P_{\perp}\geq 0$.
Since the third term on the right hand side of (\ref{19}) is positive
inside the star, the positivity of $P_{\perp}$ can be obtained by requiring
that the minimum of the two first terms in the right hand side of (\ref{19})
be positive. With a simple algebra, we get
\begin{eqnarray}
& &a\leq \frac{-243(3k^3+8k^2-\alpha)k^6}
{{(6k^3-8k^2+\alpha)}^2(3k^3+2k^2-\alpha)
(3k^3-10k^2-\alpha)}=a^{\star},\nonumber\\
& &\alpha=k^2\sqrt{9k^2-24k+28}.\label{20}
\end{eqnarray}
The inequality (\ref{20}) contains (\ref{18}).
It is interesting to note that expression (\ref{20}) is
monotonically decreasing in $k$ for $k>4$,
with $a\rightarrow 0$ for
$k\rightarrow\infty$. This indicates that anisotropy
corrections to the seed metric considered
are relevant for compact objects.\\ 
Regarding the 
behaviour of (\ref{19}), for $a\lesssim\frac{a^{\star}}{10}$,
$P_{\perp}$ is monotonically increasing (remember that $P_{0\perp}$
is monotonically increasing). Further, for $a\gtrsim\frac{a^{\star}}{10}$,
$P_{\perp}$ is monotonically decreasing up to a certain value
$r^{\star}$ where there is an absolute minimum
and is monotonically increasing for
$r>r^{\star}$.\\
Thus, for $a\gtrsim\frac{a^{\star}}{10}$, the term
$H(r)$ drastically changes the behaviour of $P_{\perp}$.
The condition $E-P_{\perp}\geq 0$ is satisfied in the range
(\ref{20}). For the anisotropic factor $S(r)$ (see (\ref{7}))
we get
\begin{eqnarray}
& &S(r)=\frac{3r^2M^2}{R^3(R^3-2r^2M)}+
\frac{12 a r^2}{R^9}\left(R^2-r^2\right)\left(-R^3+2r^2M\right)+\nonumber\\
& &\frac{9r^2a^2}{R^{15}}{\left(R^2-r^2\right)}^4
\left(R^3-2r^2M\right).\label{21}
\end{eqnarray}
For $a\lesssim\frac{a^{\star}}{10}$, expression (\ref{21}) is positive
and monotonically increasing inside the star. Conversely, when 
$a\gtrsim\frac{a^{\star}}{10}$, we have always $S(0)=0$ but
$S(r)$ is now negative up to a certain radius $r^{\prime}$ where
$S(r^{\prime})=0$. For $r \in (r^{\prime}, R]$, $S(r)$ becomes 
positive and monotonically increasing. Therefore, the physics of the system
changes for $a$ sufficiently large. Note that with a negative  value for $S(r)$
we have a "soft"' core, while for $S(r)$ positive we have a "hard"
core for the star. Further, a negative value for $S(r)$, vanishing at the centre
$r=0$, is typical of boson stars (see \cite{11}). For such stars
$S(r)=-J_{,r}^2/A(r)$, where $\Phi(r,t)=J(r)e^{-\imath\sigma t}$ is 
a complex scalar field satisfying the Klein-Gordon equation and
$\sigma$ is the frequency. 
Obviously, since Klein-Gordon equation must be satisfied for
boson stars,
we do not claim that the solution found is an exact solution 
for such stars, but only that solutions with a behavior 
for $S(r)$ related to that of boson stars can be obtained in our 
context. Physically, the parameter $a$ induces a phase transition for the 
matter inside the star. In fact, when $a\gtrsim\frac{a^{\star}}{10}$,
there is a region filled with usual matter, i.e. for $r\in(r^{\prime},R]$,
but the region $r\in[0,r^{\prime}]$ is filled with unusual matter with
$S(r)\leq 0$. Remember that $S(r)/r$ is a measure of the force 
exerced by the pressure.
In the next section we show that a "dynamic" can be introduced 
by means of a time dependence of the parameter $a$.
For the possibility of
existence of mixed boson-fermion stars, see 
\cite{21} and references therein.\\
Althought we have analyzed only a particular class of generating functions
$H(r)$, the reasonings above hold for more general choices,
provided that $H(r)$ isw monotonically increasing, regular and satisfies all
the appropriate conditions. For example, we can take
$H(r)=a{(r^2/R^2-1)}^{3+2n}$, with $a>0$ and $n$ a positive integer.
The physical features remain the ones depitced above, with an
expression for $a^{\star}$ similar to (\ref{20}) with
$k=\frac{3M}{R(3+3n)}$.
\section{Example 2: Tolman seed solution}
As a further example, we take the well know Tolman IV
perfect fluid 
solution \cite{22} derived in \cite{20}:
\begin{eqnarray}
& &A_{0}(r)=\frac{R^3(R^3-3R^2M+2r^2M)}
{(R^3-3R^2M+r^2M)(R^3-r^2M)},\label{22}\\
& &B_{0}(r)=1-\frac{3M}{R}+\frac{Mr^2}{R^3},\nonumber\\
& &E_{0}=\frac{3M(2R^6-9R^5M-7R^2M^2r^2+2r^4M^2+3R^3r^2M+9R^4M^2)}
{R^3{(R^3-3R^2M+2r^2M)}^2},\nonumber\\
& &P_{0}=\frac{3M^2(R^2-r^2)}{R^3(R^3-3R^2M+2r^2M)}.\nonumber
\end{eqnarray}
Regularity and energy conditions follow for (\ref{22}),
provided that $R>3M$.
Note that, since $E_{0}$ is monotonically decreasing, we can express
$r^2$ in terms of $E_{0}$. After some little algebra, we obtain
\begin{eqnarray}
& &r^2=F(E_{0}),\label{23}\\
& &F(E_{0})=\frac{12M^2R^5E_0+9R^3M^2-4E_0R^6M-21R^2M^3+R^2\beta}
{2(4E_0R^3M^2-6M^3)},\nonumber\\
& &\beta=\sqrt{3M^3(R-M)(69M^2-48ME_0R^3-21RM+16E_0R^4)}.\nonumber
\end{eqnarray}
Putting $F(E_0)$ in the expression for $P_0$, we get
$P_0=P_{0}(E_{0})$.\\
First  of all, we examine the solution $A=A_{0}, B=B_0 e^{H(r)}$.
It is interesting to note that the related expression for 
$P_r$ reads
\begin{equation}
P_r = P_0+\frac{H_{,r}}{Ar}.\label{24}
\end{equation}
Therefore, starting with a know perfect fluid metric and
by setting $H_{,r}=-ArP_0$, we obtain solutions with
vanishing radial pressure.\\
As an initial imput, we again take $H=a{(r^2/R^2-1)}^3,\;a\geq 0$.
With this choice, we get
\begin{eqnarray}
& &P_r=P_{0}(r)+\frac{6a}{QR^9}{\left(R^2-r^2\right)}^2
\left(R^3-r^2M\right)\left(R^3-3R^2M+r^2M\right),\nonumber\\
& &Q=R^3-3R^2M+2r^2M.\label{25}
\end{eqnarray}
For $R>3M$ and $\forall a\geq 0$, $P_r$ is monotonically decreasing.
Thus $E-P_r\geq 0$ for
\begin{equation}
a\leq \frac{(2RM+13\frac{M^3}{R}-11M^2)}{2(R^2+3M^2-4RM)}.\label{26}
\end{equation}
For $P_{\perp}$ we obtain
\begin{eqnarray}
& &P_{\perp}=P_{0}+\frac{3\gamma(r)a}{R^9}\frac{\left(R^2-r^2\right)}{Q^2}+
\label{27}\\
& &\frac{9r^2a^2}{R^{15}Q}{\left(R^2-r^2\right)}^4\left(R^3-r^2M\right)
\left(Mr^2-3R^2M+R^3\right),\nonumber\\
& &\gamma(r) = 40R^8r^2M-12MR^{10}+18r^8M^3-6R^9r^2+93R^4r^4m^3+\nonumber\\
& &18M^2R^9+6R^3M^2r^6+13R^5r^4M^2-76R^2M^3r^6-57R^7M^2r^2+\nonumber\\
& &2R^{11}-12R^6Mr^4-27M^3R^6r^2.\nonumber
\end{eqnarray}
By inspection of (\ref{27}), we see that $\gamma$ is monotonically decreasing and
gets its mininum (negative) value at $r=R$.
Further, the third term on the right hand side of (\ref{27}) is always
positive inside the star.
As a result, the values of $a$ for which $P_{\perp}\geq 0$ are
\begin{equation}
a\leq\frac{M^2R^6(R^3-3R^2M)}{|\gamma(R)|}=
\frac{(k-3)}{4(k-2){(k-1)}^2}=a^{\star}.
\label{28}
\end{equation}
The condition (\ref{28}) contains the condition (\ref{26}).
The expression (\ref{28}) is a monotonically decreasing 
function of $k=R/M$ for $k>\frac{11+\sqrt{17}}{4}$.
As we shall see later, stability requires $k\gtrsim 4.661$.
Consequently, according to \cite{2},
the role of anisotropies becomes more relevant for stable
compact objects.\\
Further, $E-P_{\perp}\geq 0$ and $P_{\perp}$ is monotonically decreasing 
in the range (\ref{28}).\\
For $S(r)$ we have:
\begin{eqnarray}
& &P_{\perp}-P_{r}=S(r)=\frac{3a}{Q^2 R^9}r^2\Delta\left(R^2-r^2\right)+
\label{29}\\
& &\frac{9r^2a^2}{R^{15}Q}{\left(R^2-r^2\right)}^4
\left(R^3-3R^2M+Mr^2\right)\left(R^3-Mr^2\right)\nonumber\\
& &\Delta = -4R^9+4r^4M^2R^3+57r^2M^3R^4+14M^3r^6-8r^2MR^6+\nonumber\\
& &24MR^8-54r^4M^3R^2+9r^2M^2R^5-33M^2R^7-9M^3R^6.\nonumber
\end{eqnarray}
By a Taylor expansion of (\ref{29}) near $r=0$ we get
\begin{equation}
S(r)=\frac{3ar^2[3Ra(R-3M)-4R^2+3M^2]}{R^5(R-3M)}+o(r^3).
\label{30}
\end{equation}
From (\ref{30}) we see that $S(r)\geq 0$ near $r=0$ for 
$a>\frac{(4k^2-3)}{3k(k-3)}$. Thus, thanks to (\ref{28}),
we have $S(r)\leq 0, \forall r \in [0,R]$ with 
$S(0)=S(R)=0$.  
Since $S(r)$ is negative, the star has a "soft" core.
The behaviour above for $S(r)$ is typical of the 
(hypothetical!) boson stars \cite{9,14}. To best represent such
objects we could require that  
(remember that a boson star has no  well-defined radius 
\cite{11}) $S(r)$ vanishes exponentially outside the star reading
a maximum negative value near $r=R$. This can be obtained, for example, 
by dropping condition (\ref{11}) and by posing
\begin{equation}
H(r)=-a{(r^2/R^2-1)}^2e^{-\frac{r^n}{R^n}}\;\;,\;\;n\geq 2\;,\;a>0.
\label{hill}
\end{equation}
However, the behaviour of $P_r, P_{\perp}$ is very similar to the one 
depicted above.\\
Concerning the equation of state of our generated solution
(\ref{25})-(\ref{27}), thanks to
(\ref{23}), we have $P_r=P_r(E), P_{\perp}=P_{\perp}(E)$,
with $P_{r,E}\in [0,1), P_{\perp,E}\in [0,1)$, where
\begin{equation}
E \in \left[\frac{6M(R-2M)}{R^3(R-M)},\frac{3M(2R-3M)}{R^3(R-3M)}\right],
\label{pa}
\end{equation}
since $E$ is monotonically decreasing.\\
For example, for $k=10, a=a^{\star}, {(P_{r,E})}_{max}\simeq 0.33$.
Therefore, the sound velocity of the fluid remains always less than 1.
For the seed solution, ${(P_{0,E})}_{max}\simeq 0.24$.
However, note that for Tolman IV solution, $P_{0,E}$ is 
a monotonically decreasing function of $E$. 
Conversely, with respect to the generated solution,
$P_{r,E}$ is a monotonically decreasing function of $E$
for (approximately)
$a \in [0,\frac{a^{\star}}{2})$, while is 
a monotonically increasing function for    
$a \in [\frac{a^{\star}}{2},a^{\star}]$.
Note that, since $E$ is a monotonically decreasing
function of $r$, we expect, on general grounds, that the sound velocity
decreases for increasing $r$. This desiderable behaviour is
obtained in our solution in the regime where anisotropies
are most relevant. Therefore, for $a \in [\frac{a^{\star}}{2},a^{\star}]$
our generated solutions satisfy all the fundamental features of 
physicity requested in the literature.\\    
For the redshift we have
\begin{equation}
z=\frac{e^{-\frac{H}{2}}}{\sqrt{B_{0}}}-1.
\label{31}
\end{equation}
With the starting choice for $H$ and 
for small $a$, we read
\begin{equation}
z=z_{0}+\frac{a}{2\sqrt{B_{0}}}{\left(1-\frac{r^2}{R^2}\right)}^3+o(a).
\label{32}
\end{equation}
The maximum redshift for our solution is attained at the centre:
\begin{equation}
z_{max}=\frac{1+\frac{a}{2}}{\sqrt{1-\frac{3M}{R}}}-1+o(a).
\label{33}
\end{equation}
If we impose the maximum redshift possible for stable boson stars
\cite{14}, i.e. $z_{max}\simeq 0.687$, for the critical 
value ${k}_{crit}$ we get
\begin{equation}
k\geq k_{crit}\;,\;k_{crit}\simeq\frac{3}{1-0.351{(1+a/2)}^2}.
\label{esc}
\end{equation}
From the reasonings belove equation (\ref{28}), the maximum value
of $a^{\star}$ is attained at $k=\frac{11+\sqrt{17}}{4}$. By putting
this value in (\ref{esc}), we obtain the estimation
$k_{crit}\simeq 4.661$, 
and thus $k>k_{crit}$
for stability. Note that the redshift is bigger than that of the 
seed Tolman IV solution.\\
By taking $B(r)\rightarrow B(r)T(t)$, the field equations
do not involve derivatives with respect to the time coordinate.
As a result, if we take $a\rightarrow a(t)$, all the expressions given 
above remain unchanged. Consequently, we can add "dynamic" to our
generated solutions by choosing a generic time dependence for $a$, 
provided that $a(t)\leq a^{\star}$. For example, we can take
\begin{equation}
a(t)=\frac{a^{\star}t^2}{{\tau}^2}\left[e^{\frac{{\tau}^2}
{{\tau}^2+t^2}}-1\right],
\label{33}
\end{equation}
$\tau$ being a constant. Physically, the term
$a(t)$ could be interpreted as the "ignition" of some
scalar field, represented by the function $H(r,t)$, at
$t=0$. This scalar field can generate a phase transition
ending with the transformation of the starting "Tolman" star
in a boson-like star. 
The considerations above suggest the 
possibility that boson stars can form by a suitable "perturbation"
of a perfect fluid (at least for the seed metric considered).
For a review for this problem see \cite{23,24}.
Finally, also for the solution depicted in section 3, we can introduce
a time dependence for $a$ given by (\ref{33}) with $a^{*}$ given by 
(\ref{20}).\\
We studied the physical properties of the generated solution in terms
of the generating functions $H(r)$ satisfying 
conditions (\ref{10})-(\ref{11b}). The results obtained cannot be
summarized as a theorem, but, almost, only necessary conditions
can be done. For example, from a lot of functions $H(r)$ considered
it seems that, in order to have $S(r)\leq 0$, $H(r)$ must be 
monotonically increasing. Conversely, in order to have $S(r)\geq 0$, 
$H(r)$ has to be chosen monotonically decreasing.\\
For example, for 
$H(r)=a{(r^2/R^2-1)}^{2n+3},H(r)=-a{(r^2/R^2-1)}^{2n+4}, a>0$,
where  $n$ is a positive integer, we have $S(r)\leq 0$ inside the star,
while the same expression with $a$ negative leads to $S(r)\geq 0$.
But monotony is a necessary but not sufficient condition.
For example, for $H(r)=a{(r^6/R^6-1)}^{2n+3}$, $S(r)$ changes sign 
inside the star.
Generally, if $H(r)$ is not monotonic inside the star, $S(r)$ changes
sign. For example, if $H=ar^2/R^2{(r^2/R^2-1)}^{2n+3}, a>0$, 
then $S(r)\geq 0$ up to a certain $r=r^{\star}$
("hard" core), while $S(r)\leq 0$ ("soft" core) 
for $r\in (r^{\star},R]$ and 
$S(0)=S(r^{\star})=S(R)=0$ 
Conversely, if $H=-ar^2/R^2{(r^2/R^2-1)}^{2n+3}, a>0$,
then $S(r)\leq 0$ up to a certain $r=r^{\star}$,
while $S(r)\geq 0$  
for $r\in (r^{\star},R]$ and $S(0)=S(r^{\star})=S(R)=0$.\\
We now study the generated solution with 
$B=B_0,A=A_{0}e^{G}$ (without loss of generality).
For example, we could take 
$G=br^2/R^2{(r^2/R^2-1)}^4, b>0$.
Also in this case, we can find a limiting value for 
$b^{\star}$ such that for $b\leq b^{\star}$ the generated solutions
satisfy all the conditions (\ref{10})-(\ref{11b}) with 
$E(r),P_r$ monotonically decreasing. 
Further, for $b\lesssim\frac{b^{\star}}{2}$,
$P_{\perp}$ is also monotonically decreasing.
For example, if $k=10$, we found $b^{\star}\simeq \frac{1}{70}$.
Further, for all the values allowed for $b$, $S(r)\geq 0$ and $S(0)=S(R)=0$.
If we put, in the example above, $b\rightarrow -b$ with $b>0$, then
we read $S(r)\leq 0$ and $S(0)=S(R)=0$ with the same range of variability
for $b$.\\
Furthermore, we have 
studied the generated solutions  by considering a lot of $G(r)$ satisfying 
conditions (\ref{10})-(\ref{11b}). 
Also in this case, necessary and sufficient conditions cannot be imposed. However,
it seems that a necessary condition to have $S(r)\leq 0$ is
that $G(r)\leq 0$, while a necessary condition to have
$S(r)\geq 0$ is that $G(r)\geq 0$.\\        
To conclude our study, we consider $A=A_0 e^G, B=B_0 e^H$.\\
The results obtained are:\\
$\bullet$ If  $H(r)$ is such that $S(r)\leq 0$ for $G(r)=0$
and $G(r)$ is such that $S(r)\geq 0$ for $H(r)=0$,
then for $a\simeq a^{\star}$ and $b\lesssim b^{\star}/n$ for a certain $n>1$ depending
on the explicit expression chosen for $(H,G)$, $H(r)$ dominates and 
$S(r)\leq 0\;\forall r\in [0,R]$ and $S(R)=S(0)=0$. Conversely, if
$a\lesssim a^{\star}/n$ and $b\simeq b^{\star}$, then $G(r)$ dominates
and therefore $S(r)\geq 0\;\forall r\in [0,R]$.\\
$\bullet$ If $H(r)$ is such that $S(r)\geq 0$ for $G(r)=0$
and $G(r)$ is such that $S(r)\leq 0$ for $H(r)=0$,
then for $a\simeq a^{\star}$, $b\lesssim b^{\star}/n$, $H(r)$
dominates and $S(r)\geq 0\;\forall r\in [0,R]$. Conversely, if
$a\lesssim a^{\star}/n$ and $b\simeq b^{\star}$, then
$G(r)$ dominates and $S(r)\leq 0\;\forall r\in [0,R]$.\\
$\bullet$ If  $H(r)$ is such that $S(r)\geq 0$ for $G(r)=0$
and $G(r)$ is such that $S(r)\geq 0$ for $H(r)=0$, then
$S(r)\geq 0\;\forall r\in [0,R]$.\\
$\bullet$ If  $H(r)$ is such that $S(r)\leq 0$ for $G(r)=0$
and $G(r)$ is such that $S(r)\leq 0$ for $H(r)=0$, then
$S(r)\leq 0\;\forall r\in [0,R]$.\\
$\bullet$ In all the other cases, $S(r)$ changes sign inside the star.\\
Obviously, the values of $a^{\star}, b^{\star}$ depend on the chosen 
generating functions $(H,G)$, but generally we get expressions similar to
(\ref{28}).\\
From the discussion above, it follows that, in particular,
configurations with $S(r)\leq 0$ can be generated also by varying the energy 
density of the seed perfect fluid solution.\\
The results obtained with Tolman IV, in relation to the possibility of building 
interior solutions with $S\leq 0$,
do not represent a "lucky" case.
The same "phenomenon" can be easily obtained with other perfect
fluid seed metrics. As a further example, we can take the seed Heint IIa
metric ( see \cite{22}). The metric functions are ($R>3M$):
\begin{eqnarray}
& &B_0 = K{\left(1+ar^2\right)}^3\;,\;
A_0 = {\left[1-\frac{3ar^2}{2}\left(\frac{1+C{(1+4ar^2)}^{-\frac{1}{2}}}
{1+ar^2}\right)\right]}^{-1},\nonumber\\
& &a=\frac{M}{R^2(3R-7M)}\;,\;K=\frac{{(3R-7M)}^3}{27R{(R-2M)}^2},\nonumber\\
& &C=\frac{\sqrt{3}}{R}(3R-8M)\sqrt{\frac{R-M}{3R-7M}}.\label{H1}
\end{eqnarray}
With the metric (\ref{H1}), we have $S(r)\leq 0\;\forall r\in [0,R]$
with $H(r)$ given by (\ref{16}) (with $a>0$) and $S(0)=S(R)=0$,
privided that:
\begin{equation}
a< \frac{24k^2-65k+\sqrt{\frac{3(k-1)}{3k-7}}(9k-24)}
{6k(3k-7)}.\label{H2}
\end{equation}
By using the less restrictive condition (\ref{28}), it is follows that,
for $a>0$, in adjoint, our generated solutions satisfy all
energy conditions and $\forall r\in [0,R]$ are regular and
$E(r), P_r(r), P_{\perp}(r)$ are positive and monotonically decreasing.
Thus the seed metric Heint IIa can also be used to build models
recalling boson stars. Further, the procedure above
for building boson-like stars can be
successfully  applied to Durg IV and Durg V seed metrics with
$H(r)$ given, for example, by (\ref{16}).
(see\cite{22}).

\section{Example three: Building gravastars solutions}
Gravastars (see \cite{25,26}) represent an alternative to black holes.
These are described in terms of an interior with the equation of state
$E=-P$ (de Sitter) matched with a thin shell with $E=P$.
In \cite{27} a more generalized class of gravastars is discussed for which
no thin shell is  needed. The matter is composed by an anisotropic
fluid with a continuous expression for $P_r$ with the following 
boundary conditions (see \cite{27}):\\
1) $\forall r\in [0,R],\;E\geq 0$\\
2) $E(0)=-P_r(0)=-P_{\perp}(0)$ (only at the centre), 
$P_{r,r}(0)=0$.\\
3) For  $r^{*}\simeq 2M$, there exists a positive maximum for $P_r$.
For $r\in [r^{*},R)$, $P_r$ is monotonically decreasing and the
stars is filled with "ordinary" matter. For
$r\in [0,r^{*})$, $P_r$ is monotonically increasing. Further,
$S\geq 0$.\\
4) There exist exactly two points at which $P_r=0$.\\
All these conditions can be easily fulfilled.
As an example, we consider again the Tolman IV solution
together with
\begin{equation}
A=A_0,\;H(r)=a{\left(1-\frac{r^2}{b^2}\right)}^3,
\;a>0,\;b\in[0,R).
\label{hh1}
\end{equation}
In this way, the solution for $r\in[0,b)$ is composed of anisotropic 
matter, while for $r\in[b,R]$ the matter content is that given by
Tolman IV solution (\ref{22}) and the two metrics are smoothly
joined at $r=b$.
Condition 1) is obviously satisfied. Conditions 2) follow
by setting:
\begin{equation}
a=\frac{Mb^2(R-M)}{R^3(R-3M)}.
\label{H3}
\end{equation}
Regarding conditions 3), we must to chose $b\simeq 2M$.
It follows that $S\geq 0$ and $S(0)=S(b)=0$. Finally, the behaviour
for $P_r$ is the one required by the conditions  3),4). For example, by
setting $b=2M, R=10M$, we found $r^{*}=1.9986M<b$.
The graph of $P_{\perp}$ is similar to the one for $P_r$ and
for $P_r$ and $P_{\perp}$ follow expressions similar to those
given by (\ref{25}), (\ref{27}). 
Obviously, at $r=R$ our solution is smoothly matched to the
vacuum Schwarzschild solution.
With the simple procedure described in this section, specific
workable models for gravastars can be built simply by
setting an appropriate perfect fluid solution in the region
$r\in [b,R]$.\\
We can use other seed metrics than
Tolman IV. As a further example, we can use Heint IIa metric given by
(\ref{H1}). For this seed metric, it is easy to see that all the conditions
$1-4$ can be fulfilled with (\ref{hh1}), provided that, instead of
(\ref{H3}), we take: 
\begin{equation}
a=\frac{Mb^2\left[3R+\sqrt{\frac{3(R-M)}{3R-7M}}(3R-8M)\right]}
{2R^3(3R-7M)},\label{HH4}
\end{equation}
with $b\simeq 2M$.
We do not enter in the details, but only mention the fact that
the procedure above to build gravastar-like models can also be obtained,
for example, with the Durg IV and Durg V seed metrics, with respectively
\begin{eqnarray}
& &a=\frac{4M b^2}{21 R^3{(4R-9M)}^2}
\left[q\sqrt{\frac{4(R-M)}{4R-9M}}+52R^2-117 MR\right],\label{BB1}\\
& &a=\frac{5M b^2}{336R^3{(5R-11M)}^3}\left[s
{\left(\frac{5(R-M)}{5R-11M}\right)}^{\frac{1}{3}}+29161M^2R-26510MR^2
+6025R^3\right],\nonumber\\
& &q=224M^2-172MR+32R^2,\nonumber\\
& &s=2375R^3+52425M^2 R-44800M^3-19800MR^2.\nonumber
\end{eqnarray}
With the relations (\ref{BB1}), also for Durg IV, V 
all the conditions $1-4$ can be sasfied together with $b\simeq 2M$.
\section{Conclusions}
Motivated by the increasing interest in anisotropic fluids, we 
presented a simple procedure for building physically reasonable spacetimes
with local anisotropies starting from a known regular physically sensible
seed metric. The principal aim of this paper has been the construction
of a simple algorithm for exploring the physical consequence derived by
a departure from a given configuration by means of given parameters.
We studied in detail two physically interesting cases by taking
as seed spacetimes the Florides solution without radial pressure and 
with constant energy-density and the Tolman  IV perfect fluid solution. 
In particular, we showed that spacetimes with a behaviour of the 
anisotropy factor $S(r)$ closely related to that of boson stars
can be generated starting from a suitable perfect fluid solution (at least with
the  seed solutions considered!). A time dependence can be 
introduced in the $B(r)$ metric coefficient simply by taking
$a\rightarrow a(t)$. The same is not possible for $A(r)$ because,
in this case, $G_{rt}\neq 0$ and a heat flow term appears. 
Physically, the term $a(t)$ could be 
interpreted as the "ignition" at $t=0$
of some scalar field represented by
the term $H(r)$ in the metric component $g_{tt}$.
Further,
starting with an appropriate seed perfect fluid metric,
also a gravastar-like model can be built with the desired boundary
conditions.\\  
In particular, we have obtained gravastar-like models and boson-like
stars by using the following seed metrcics: Tolman IV,
Heint IIa, Durg IV, Durg V.\\  
Finally, we point out that the technique used in this paper
can also be used by taking a generic starting metric representing
a regular physically reasonable perfect fluid with an
oculate choice for the functions $H(r),G(r)$.

\end{document}